 \documentclass[final,1p,times]{elsarticle}
\usepackage{graphicx}
\usepackage{amssymb}
\journal{Extreme Mechanics Letters}

\begin{document}

\begin{frontmatter}


\title{Elastic stabilisation of wrinkles in thin films by auxetic microstructure}



\author[1,2]{Alessandra Bonfanti}
\author[2]{Atul Bhaskar} 

%\affil[a]{Department of Engineering, University of Cambridge, Cambridge, UK}
%\affil[b]{Faculty of Engineering and Physical Sciences, Southampton Innovation Boldrewood Campus, University of Southampton, SO16 7QF, Southampton, UK}

\address[1]{Department of Engineering, University of Cambridge, Cambridge, UK, ab2425@cam.ac.uk}
\address[2]{Faculty of Engineering and Physical Sciences, Southampton Innovation Boldrewood Campus, University of Southampton, SO16 7QF, Southampton, UK}

\begin{abstract}
Thin elastic sheets and membranes are known to wrinkle when they are stretched--the associated physics is highly non-linear. The mechanics of thin films that exhibit unusual behavior upon stretching, when they possess auxetic structure, i.e. when their apparent Poisson's ratio is negative, is presented here. Wrinkling is now suppressed within the bulk of auxetic films when tensioned, whereas localized creases confined to the clamps, that decay away exponentially, appear. 
These edge wrinkles are characterized for their amplitude and wavelength experimentally, theoretically, and computationally, which show excellent agreement with expected trends. The scaling for amplitude, wavelength and decay rate upon film properties and tension is obtained using simple analyses based on kinematic mismatch resulting from lateral Poisson's expansion.
\end{abstract}


\end{frontmatter}

%\linenumbers

Natural as well as engineered materials frequently appear in the form of membranes, films and elastic sheets. Synthetic skin \cite{huck2005artificial,efimenko2005nested} and biological membranes provide partitioning and containment for soft matter where wrinkling is of practical concern; stiff films mounted over soft substrates are common in many electronic devices \cite{kim2010waterproof}; and membranes are often used within shape morphing aerodynamic surfaces--all with significance to a wide range of natural phenomena such as morphogenesis \cite{wang2015three,ball1999self,thompson1942growth}, cell motility \cite{harris1980silicone}, tissue differentiation \cite{guvendiren2010control}, as well as to technological applications such as microfabrication \cite{ahmed2010high,okayasu2004spontaneous}, adaptive aerodynamic drag control \cite{terwagne2014smart}, metrology of thin films \cite{stafford2004buckling,huang2007capillary}, etc. Thin rectangular films develop wrinkles when stretched, except near clamping edges where they remain taut \cite{cerda2003geometry}. Here we report the converse phenomenon for structured thin films with apparent negative Poisson's ratio. Such films remain wrinkle-free within the bulk of a membrane when stretched, except near the clamped edges. Further, we develop a simple kinematic model to characterize the amplitude and the wavelength of anomalous wrinkles confined to the clamping edges. %that are interpreted here as elastic instability analogues of the well-known phenomenon of trapped edge waves. 
Wrinkle-free stretching, as reported here, is found to be robust in simulations as well as experiments and consistent with our mathematical analysis. 

The possibility of fabricating architectured materials has enabled the development of modern meta-materials with unusual properties that are not commonly found in naturally occurring materials. In the realm of elastic metamaterials, apparent properties leading to negative refractive index for sound propagation, negative Poisson's ratio (such material being termed as auxetic) \cite{lakes1987foam}, negative compressibility \cite{baughman1998materials} and negative coefficient of expansion \cite{mary1996negative} have been observed before. The use of stretchable planar devices that adapt to the shape of natural soft surfaces to monitor tissue response has been suggested  \cite{viventi2011flexible,kim2011epidermal}. Such membrane devices can develop wrinkles when compressive stresses are generated \cite{kim2011hierarchical}. Buckling of thin films under compression is a well-known problem in mechanics \cite{bowden1998spontaneous,vandeparre2011wrinkling} and practical solutions to arrest this include increasing film thickness, or inclusion of nanoparticles to lower the internal compressive stress \cite{hendricks2007wrinkle}. Here we explore elastic stabilization of lattice membranes by tailoring the apparent Poisson's ratio, a property bound theoretically within the narrow range $-1<\nu<+0.5$ for isotropic materials, and is positive between $0.25-0.35$ for most naturally occurring materials. By apparent properties, we mean the effective properties of micro-structured film material that would have the same bulk response as that of porosity-filled film of the same shape and size.    

Because of their low bending stiffness, membranes are prone to wrinkling and structural instabilities such as folding and creasing. When an elastic sheet is compressed in its plane, there is a competition between the strain energy stored within the elastic solids and the potential energy of external loading, the latter being associated with the so-called ``kinematic stiffness". Wrinkling of membranes under in-plane shear has also been studied \cite{wong2006wrinkled}, the state of stress being compressive at 45 degree orientation to shear. As opposed to these, Cerda and Mahadevan \cite{cerda2003geometry} reported the formation of wrinkles within the bulk of rectangular strips of thin films, when they are stretched, except near the clamped edges where they remain taut. This unusual elastic instability is under no compressive stresses across the waviness of wrinkles and is best explained via the structure requiring geometric fit to the same size as what the Poisson's shrinkage would produce. The essence of Cerda-Mahadevan-type instability, therefore, is the kinematic constraint posed by Poisson-contraction.

\section{Experimental results} 
\begin{figure}[h]
\centering
\includegraphics[width=8.5cm]{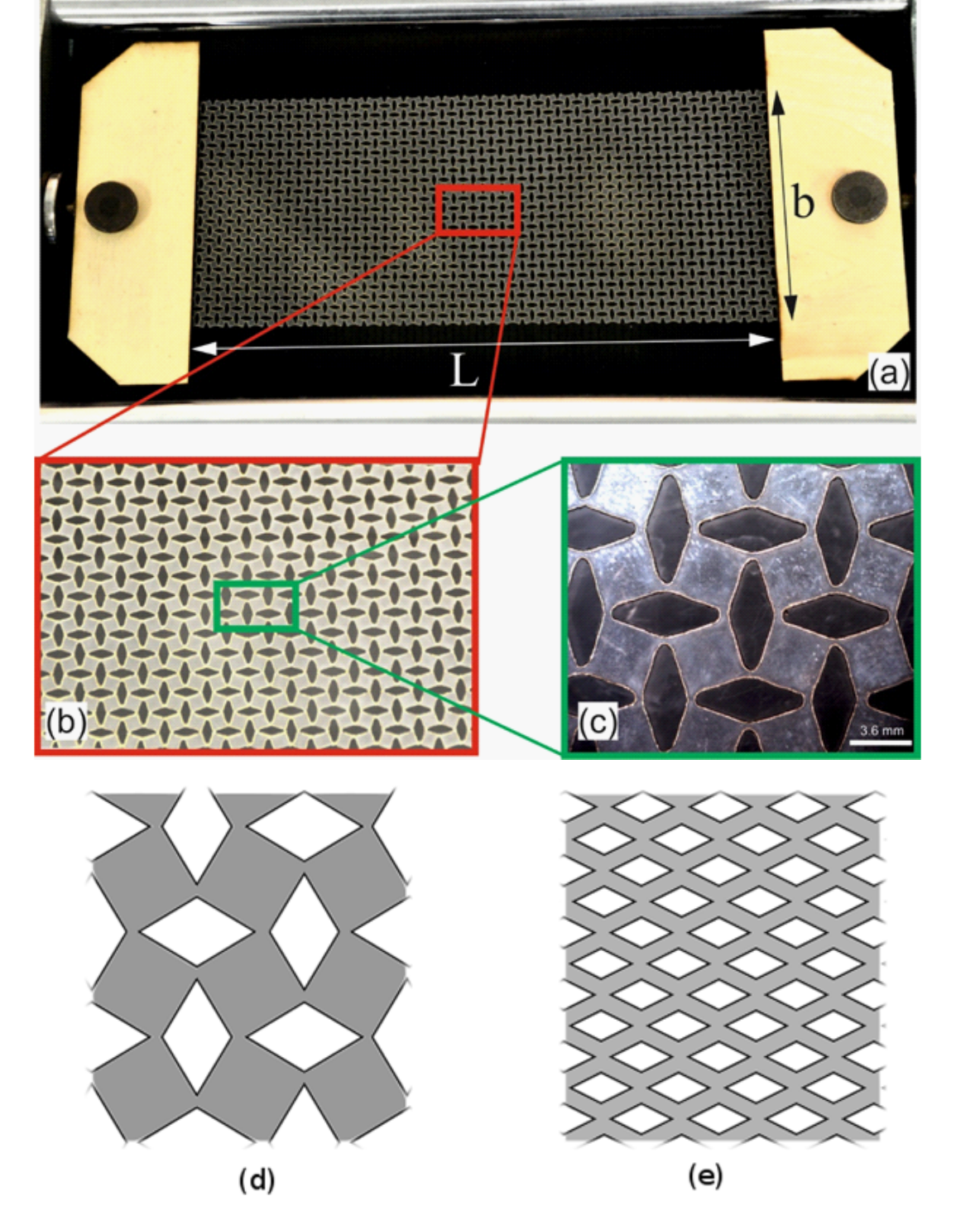}
\caption{Structured cellulose acetate film. (a) Complete experimental sample with perforation, clamped at the ends. (b) enlarged view showing the lattice of diamond shaped periodic holes. (c) A closer look highlights the square elements connected at small ligaments that allow rotation increasing the volume of the voids when stretched. Increase in void upon stretching leads to apparent negative Poisson's ratio of $\nu \approx -1$. (d,e) Sketches of the auxetic and non-auxetic geometries, respectively. }\label{fig:fig_1}
\end{figure}

Wrinkling of sheets in tension due to Poisson's shrinkage suggests that it could be suppressed within their bulk, if we could engineer the architecture of the material so that the {\em apparent} Poisson's ratio of the structured film  is {\em negative}. There are several well-known auxetic microstructures. Here we consider one such lattice geometry (Figure \ref{fig:fig_1} (a)-(d)); the auxetic behavior achieved by the rotation of the squares leading to the enlargement of the voids upon stretching \cite{gatt2015hierarchical}. We laser-cut perforated samples of length $L$, and width $b$, $L>>b$, Figure \ref{fig:fig_1} (a) from acetate sheets (zoomed micro-structure of the sample in Figure \ref{fig:fig_1} (c)), where squares of side $\sim4$ mm are connected, leaving diamond shaped holes to form the lattice, schematic sketch  in Figure \ref{fig:fig_1} (d)). By contrast, a perforated film with diamond shaped voids in Figure \ref{fig:fig_1} (e) possesses positive apparent Poisson's ratio. The choice of the lattice geometries is guided by the ease of manufacture as well as minimal degree of perforation of continuous film required to achieve significant auxetic behaviour. Other auxetic lattices are expected to behave similarly, but not considered here. Here we refer to the direction along the length of the rectangular film as {\em longitudinal}, along the width as {\em lateral}, and perpendicular to the plane of the film as {\em transverse}. 

\begin{figure}[h]
\centering
\includegraphics[width=12cm]{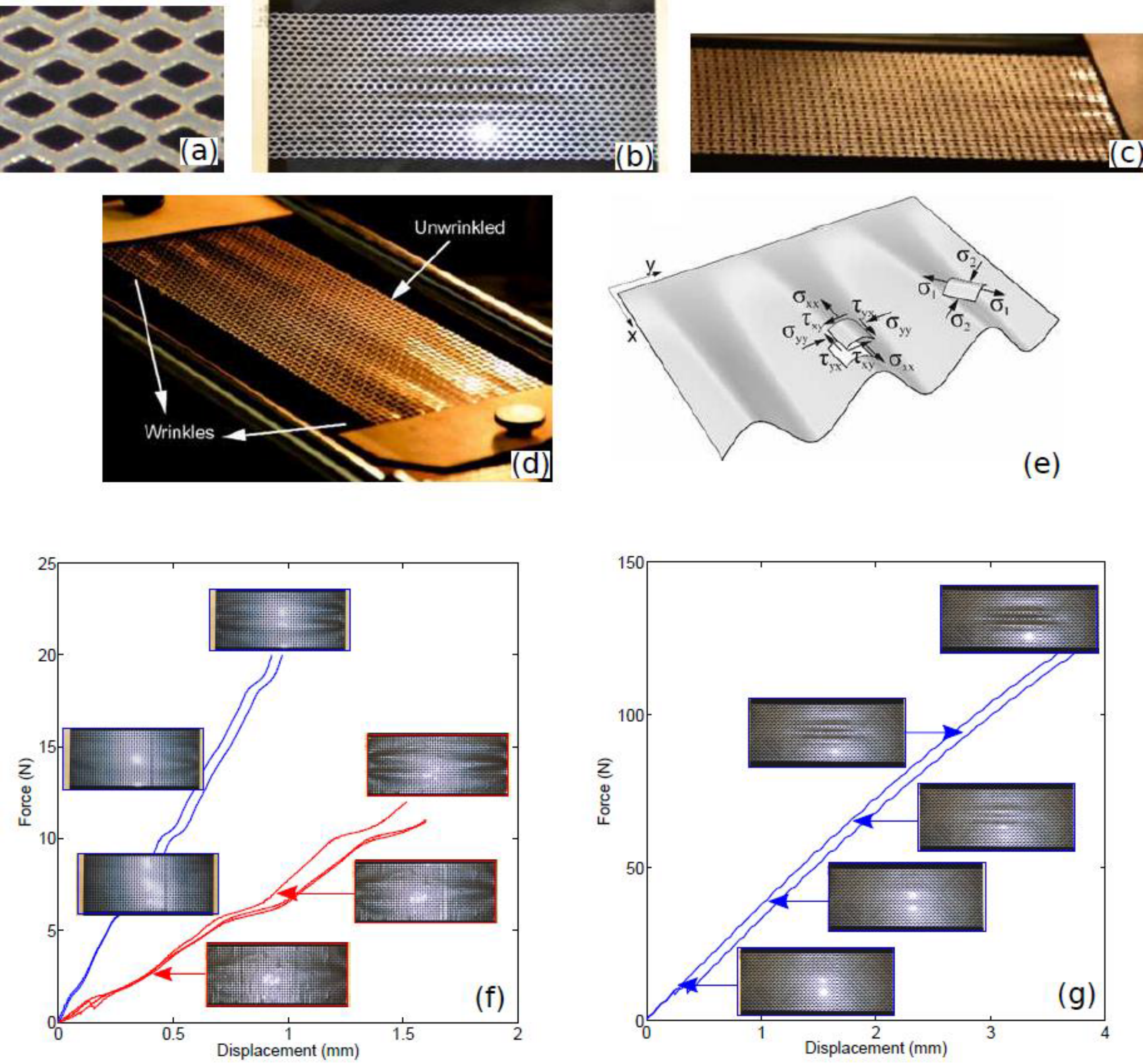}
\caption{Microstructured films under tension. (a) Zoomed image of the microstructure associated with positive Poisson's ratio, (b) structured film with positive apparent Poisson's ratio shows wrinkling at the centre except at the clamped edges, (c) Perforated film (microstructure in Figure \ref{fig:fig_1} (a)-(d)) with apparent negative Poisson's ratio is stabilised at the centre – only half of the film is shown. Anomalous localised wrinkles at the edges are observed leaving the bulk of the centre of the film to be wrinkle-free. (d) The amplitude of the anomalous edge folds increases upon increasing tension and spreads towards the centre of the rectangular perforated film, (e) A schematic diagram of the film near the clamped edges that shows the development of a biaxial state of stress leading to folds, (f) development of localised wrinkles at the edges upon increase tensile load for two films with auxetic porous structures; the x-axis is the displacement in the plane of the film along the direction of tension (y-axis). The amplitude of out-of-plane edge wrinkles increases with increasing tension (g) Wrinkling of non-auxetic perforated film (microstructure in Figure \ref{fig:fig_1} (e)) upon stretching, except at the edges that are fixed.}\label{fig:fig_2}
\end{figure}

First, we applied tensile stress in the displacement controlled mode of tensometer (see Methods) to a perforated rectangular structured film of lattice geometry shown in Figure \ref{fig:fig_1} (e). This lattice possesses positive apparent Poisson's ratio. Upon stretching, the lattice film (Figure \ref{fig:fig_2}(a)) developed wrinkles at the center (Figure \ref{fig:fig_2}(b), which shows wrinkling within the bulk), except at the clamped edges where it is taut. This behavior is analogous to one reported by Cerda et al.\cite{cerda2002thin} previously for imperforate homogeneous films made of material with positive Poisson's ratio. When the diamond shaped holes are oriented on a 2D lattice differently (Figure \ref{fig:fig_1} (a)-(d)), the apparent Poisson's ratio of the film becomes negative; the film material itself always has positive Poisson's ratio. When an auxetic film is pulled longitudinally, it remains wrinkle-free at the centre (left end of Figure \ref{fig:fig_2}(c) which shows only half of the complete film). Small amplitude anomalous wrinkles appear at the clamping locations (right end of Figure \ref{fig:fig_2}(c)). This is in clear contrast with the wrinkling behavior of thin films made of material with positive Poisson's ratio as reported in \cite{cerda2003geometry,cerda2002thin}.  Rectangular films of finite length and width will eventually be filled with wrinkles when tension is increased – for films of both positive and negative apparent Poisson's ratio (Figure \ref{fig:fig_2}(d)). The important distinction, however, is that the bulk of auxetic film, in a long and thin strip, will be taut whereas the opposite happens for a non-auxetic film in which the taut region is localized at the clamping edges.   

Consider the state of stress on an elastic element close to the clamped edge (Figure \ref{fig:fig_2}(e)), which is biaxial due to fixed width imposed at the clamps. Consider a tensioned film that is free along the long edges, and is laterally fixed at short edges. Lateral contraction would occur for positive Poisson's ratio membranes but dilation for auxetic films. Because of lateral inextensibility at the clamps, lateral stresses near the clamps need being superposed on the remote longitudinal stresses, resulting in biaxial state of stress (and strain). Apparent stresses in auxetic films can be transformed in the principal directions by rotation of the coordinates such that shear stress vanishes—principal stress $\sigma_1$ along the anomalous wrinkles is tensile, whereas $\sigma_2$ across them is compressive, the latter causing folds to form. This is in contrast with the films made of positive Poisson's ratio material (or lattice films with positive apparent Poisson's ratio) tensioned similarly, where the principal stresses are both tensile within the biaxially stressed clamping region which keeps this small region taut. In many ways, the situation is mirrored in auxetic films – the bulk of the auxetic film at the center is wrinkle-free except the localized small regions near the clamps. Upon increasing tension excessively, the end effects penetrate into the centre of the films for auxetic and non-auxetic structured films, as shown on the force-stretch diagram (Figure \ref{fig:fig_2}(f) and (g) respectively) – auxetic films show stabilized centre, except anomalous wrinkles at the edges; the opposite is observed for non-auxetic films. Six samples (four auxetic in Figure \ref{fig:fig_2}(f) and two non-auxetic in Figure \ref{fig:fig_2}(g)) were pulled in tension (Methods). The images have been placed at the corresponding locations on the force-stretch plane. The slopes the two pairs of curves in Figure \ref{fig:fig_2}(f) (in blue and red respectively) represent two different lattices with the same shape but different geometry resulting in different stiffness. The three pairs of approximate straight lines represent three nominally identical sets, the differences within each pair is attributed to experimental errors and geometrical/material inconsistencies across different samples. 

%\section{Confined wrinkles as trapped waves}
%Spatially localized wrinkles confined to the clamped edge of a stretched auxetic membrane, as reported here, have an interesting physical analogy. A host of wave phenomena show edge waves or localized mode, perhaps the earliest known example being Rayleigh surface wave that propagates on the surface freely but decays depth-wise exponentially. Trapped shallow water waves travel along the shore but are evanescent away from it \cite{ursell1952edge}. This behavior is characterized by a sinusoidal dependence along the edge, which is the direction of propagation and an exponential decay away—-the direction along which the wave is evanescent or confined (see, e.g. \cite{kaplunov2000free} for elastic waves). Thin plates possess such spatially localized wave propagation modes known as Konenkov waves. Here we report an elastic instability mode that is ``trapped'' longitudinally but ``propagates” laterally; i.e. ripples are observed along the edge but decay away from the edge. Such trapped instability modes in wrinkling do not seem to have been reported before. Associated with this trapped instability mode, is a length scale over which the wrinkles decay. This is mirrored for non-auxetic films where there is a small persistence length near clamping over which the film is taut.   

Auxetic membranes expand laterally upon stretching, except at the edges where the film is restrained. This poses kinematic requirements that is just the opposite of that for stretched non-auxetic membranes which need to fold within the bulk, in order to fit the width as dictated by Poisson's shrinking. Thus, we obtain a largely wrinkle-free center for a long auxetic film stretched along its length, except at the clamped edges where the state of stress is biaxial. ``Blister-like” localized instability, which decays away from the clamps but ripples along the transversely and laterally fixed edge, is observed experimentally (the wrinkled film close to the clamps in Figure \ref{fig:fig_2} (d); the tension was increased well beyond the onset of instability to visually highlight the shape of the well-developed wrinkles). In our experimental set up here, the extremities are forced to stay transversely flat, so that edges are clamped laterally as well as transversely, so that edge wrinkles that are flattened forming blisters that exponentially decay away from the fixed edges.

\begin{figure}[h]
\centering
\includegraphics[width=10cm]{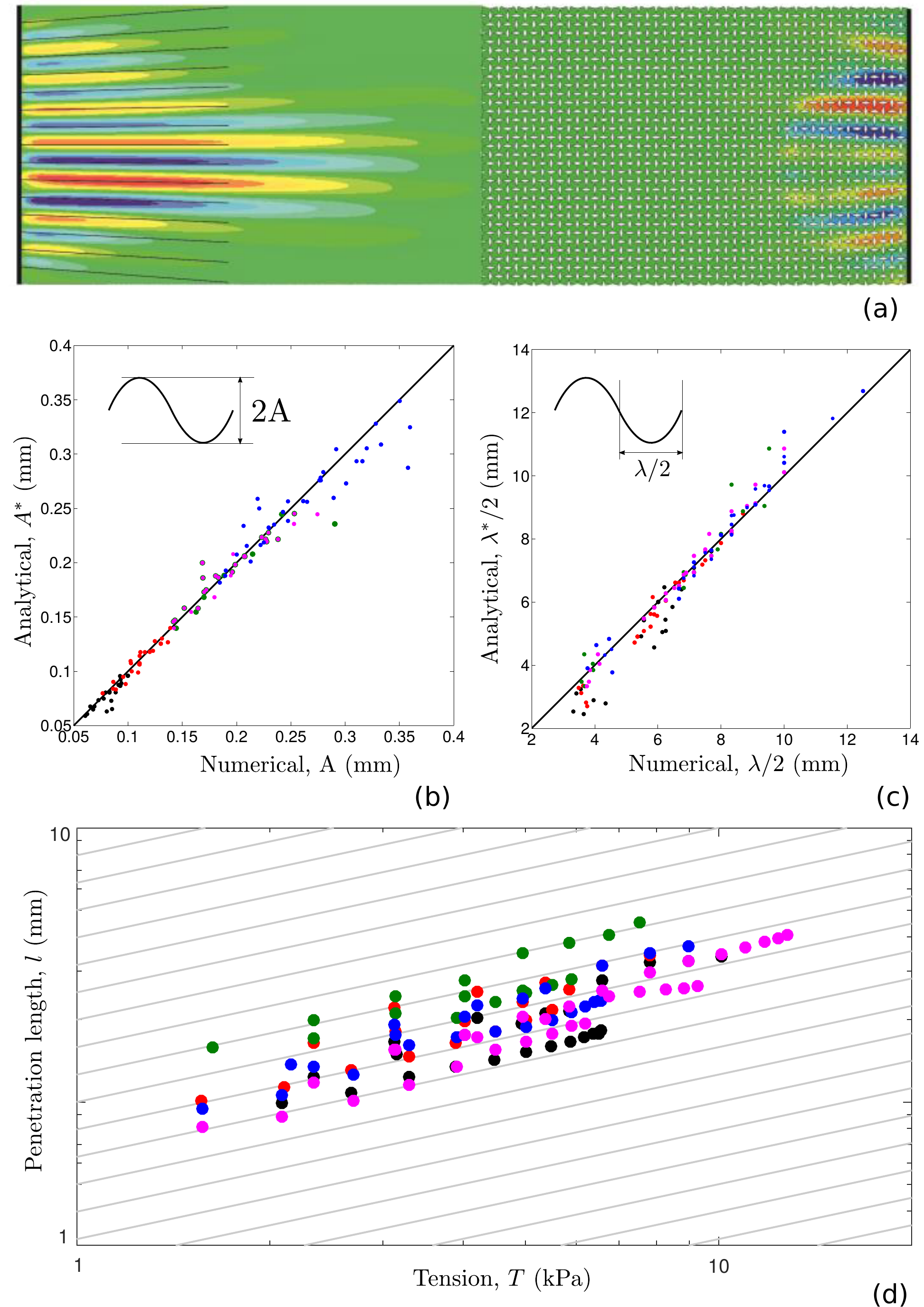}
\caption{Anomalous edge wrinkles: comparison between theoretical predictions based on the kinematic model and simulation. (a) Homogeneous imperforate film with negative effective Poisson's ratio material (left half of the figure), and perforated film with auxetic microstructure of Figure \ref{fig:fig_1} (a)-(d) (right half of the figure). The principal stress directions are plotted using black lines superposed on top of the numerically calculated out-of-plane displacements shown using colored contours. Note that the black lines are well aligned with the theoretically calculated principal directions of the biaxial stress (b) Amplitude of edge-localized wrinkles $A^*$ as given by the kinematic analysis above vs that observed in simulations on thin films with auxetic material properties. Each point refers to a case of Poisson's ratio, film thickness $t$ and $L/b$ ratio of the rectangular planform (c) the same as in Figure b, but for analytically calculated wavelength of the localized wrinkles vs finite element simulations (d) The penetration length of the wrinkles scales with the applied tension to the power of $1/2$ as shown in the log-log graph. }\label{fig:fig_3}
\end{figure}

For the ease of analysis that could yield to a meaningful understanding of the stabilization as well as one that would enable quantitative characterization of anomalous edge wrinkling, consider a simplified clamping condition such that the film is constrained laterally (thus creating lateral Poisson mismatch when tensioned) but is free normal to the plane of the film. Although difficult to realize this in practice, the condition could be simulated easily in computational experiments. This simplification facilitates analytical treatment and gives insight into the phenomenon while deliberately compromising on realism of the transverse clamping condition but respecting the lateral clamping.  Dependence of the amplitude and the wavelength of edge wrinkles upon the remote stress is theoretically characterized and computationally verified next.      

\section{Numerical experiments and a kinematic theory of edge wrinkles}

Strain along the longitudinally applied tensile stress $\sigma$ is given by $\epsilon^{||}=\sigma/E$, where $E$ is the apparent Young's modulus of the film. The lateral strain, far from the clamped edges, is {\em extensional} and is given by $\epsilon^{\perp}=-\nu\epsilon^{||}$, where $\nu<0$ is the apparent Poisson's ratio of the perforated auxetic film. Wrinkles can be treated as spatially localized instability modes and be represented as sinusoidal ripples in the $y$-direction that exponentially decay in the $x$-direction, when the constraint at the ends is lateral but not transverse. Such a length-wise decaying wrinkling pattern can be represented by $w(x,y)=A \sin(\pi y/\lambda) \mathrm{e}^{-\pi x/l}$; where $l$ is the associated decay length scale. The total lateral dilatation at the length-wise center of the film is the sum of the increase in lateral projections, which equals the integral over width $b$ of the square of the $y$-wise slope, and can be evaluated analytically  
\begin{equation}
\Delta b = 	\frac{1}{2} \int \left( \frac{\partial w}{\partial y} \right)^2 dy = \frac{n^2 \pi^2 A}{b}
\end{equation}
This provides us with the dependence of the lateral expansion upon the amplitude and the number of wrinkles, $n=b/\lambda$. Combining this with lateral Poisson extension far from the clamped edges, $\Delta b=\nu b\sigma/E$, we obtain the ratio of the amplitude of wrinkles to width,  $A^*=A/b$, in terms of the applied non-dimensional tensile stress scaled with respect to the Young's modulus $\sigma^*=\sigma/E$, as
\begin{equation}
A^* = 	\frac{(|\nu| \sigma^*)^{1/2})}{n \pi} 
\end{equation}
The wavelength of the wrinkles can be estimated from the vertical equilibrium of a shell \cite{wong2006wrinkled} which requires $\sigma_1 \kappa+\sigma_c \kappa ^{\prime}=0$, where $\kappa=\partial ^2w/\partial x^2$ and $\kappa ^{\prime}=\partial ^2 w/\partial y^2$ are the $x$-wise and $y$-wise curvatures respectively; $\sigma_c=\pi^2 Et^2/[12(1-\nu^2 ) \lambda^2]$ is the critical buckling load for a simply supported plate strip of thickness $t$, and $\sigma_1$ approximately equals the applied stress $\sigma$. Inserting the expressions for curvatures after differentiating $w(x,y)$ twice, we obtain an expression for the wavelength of the edge-localized wrinkles as
\begin{equation}
\lambda^* = \frac{(	\pi t l)^{1/2})}{[12(1-\nu^2 ) \epsilon^{||} ]^{1/4}} 
\end{equation}
Further refinement in the theoretical characterization of the anomalous edge-wrinkles is possible by recognizing that the constraint at the clamped edges sets up a biaxial stress field (Figure \ref{fig:fig_2} (e)): tensile stresses $\sigma_x$ longitudinally, compressive stress $\sigma_y$ laterally, and shear stress $\tau_{xy}$  due to kinematic mismatch caused by the lateral Poisson's expansion away from the edge but resisted laterally by the clamp.  The state of stress is fairly complex and can be calculated by adapting the analysis provided by Bentham \cite{benthem1963laplace} to negative Poisson's ratio. Moreover, localized wrinkles at the edges are not perfectly parallel to the stretch direction due to the non-uniformity of the lateral stress across the width, causing slight tilt of the wrinkles that varies laterally. Our analysis enables us to quantitatively estimate this lateral tilt of the edge-localized wrinkles. Once the components of the direct stress $\sigma_x$, $\sigma_y$ and the shear stress $\tau_{xy}$  are calculated from the adaptation of Bentham's analysis, the principal stresses $\sigma_1$,$\sigma_2$ and the principal directions (using Mohr's circle, or formally eigenvector of the matrix of the components of the stress tensor) were obtained. Principal tensile direction indicates the direction of the localized wrinkles. The relatively small compressive principal stress is across the wrinkles. While our simple analysis is based 
on kinematic compatibility necessitated by the Poisson-effect, mathematically rigorous approaches such as tension field theory \cite{steigmann1990tension} have been used when large stresses are along wrinkles but negligible stresses across them. Theoretically calculated direction of wrinkles are compared with those of  wrinkles observed in finite element simulations (Methods) next.  

We carried out simulations for stretching of imperforate continuum films made of negative Poisson's ratio material using Finite Element Analysis (FEA, see Methods) that show wrinkle-free bulk except near the edges. The direction of edge wrinkles aligns well with the mathematically derived directions of principal stresses--black arrows at the left clamps are shown on the left half of Figure \ref{fig:fig_3} (a); pattern of localized wrinkles at the right clamped end is symmetric and is omitted.  There is a remarkable consistency between the predicted directions of localized folds--titled slightly due to biaxial nature of the stress field near the clamped edges--and those observed in FEA simulations. Finally, computational experiments using FEA were carried out for micro-structured auxetic film under tension. Wrinkles confined to the right clamped edge for such perforated auxetic films under tension as obtained in simulations are shown on the right half of Figure \ref{fig:fig_3} (a); the left half has symmetric patterns of mechanical response and is omitted.  Tension was progressively increased during simulations ensuring clear visual appearance of wrinkles, allowing them to spread towards the center. The behavior of continuum as well as perforated films with negative effective Poisson's ratio--that the center of the film remains largely taut--are consistent with our laboratory experiments and demonstrate that the phenomenon is robust which can be attributed to the lateral expansion of auxetic films.  

 Wrinkles at the clamped edges can be characterized systematically in simulations that enable us to study a wide range of materials, loading and geometry. The amplitude (Figure \ref{fig:fig_3} (b)) and the wavelength (Figure \ref{fig:fig_3} (c)) were measured from the numerical results for films with a range of properties and compared against the predictions from our simple kinematic theory showing excellent agreement between the two; the solid line equally inclined to the two axes corresponds to perfect agreement. Each dot represents measurements of amplitude of the localized and wavelength from a computer experiment. Different colors of the markers refer to different values of the apparent Poisson's ratio ($-0.2,-0.4,-0.6,-0.8,-0.9$ respectively). For each Poisson's ratio, films with different aspect ratio $L/b$ (10, 7 and 5) and thickness $t$ (0.01 mm, 0.05 mm and 0.1 mm) were stretched in simulations. The decay length scale $l$ was taken as half of the length of the film, i.e. $L/2$ for order of magnitude estimates. Over this length scale, the amplitude decays to a value of $\mathrm{e}^{-\pi}\approx 4.3\%$ of its amplitude at the edge. 

 \section{Decay length scale and its scaling with tension}

Carrying out a 1-D analysis--in the spirit of Cerda-Mahadevan \cite{cerda2003geometry}--the local wrinkling behavior at the edges of auxetic films is governed by $B(\partial^{4}w/\partial y^{4})- T(\partial^{2}w/\partial x^{2})-b(\partial^{2}w/\partial y^{2})=0$. Seeking a separable solution of the form, $w(x,y)=\sin ky X(x)$, we obtain the length-wise variation of edge-localized creases given by
\begin{equation}
\frac{{\rm d}^{2}X}{{\rm d}x^{2}}-\alpha^{2}X=0, \quad {\rm where} \quad \alpha^{2}=\frac{k^{4}B+k^{2}b}{T}.
\end{equation}
Therefore, $x$-wise dependence of the wrinkling amplitude is given by $X(x)\sim e^{-\pi x/l}$, which shows an exponential decay, if $b$ is taken as a constant, suggesting that edge wrinkles are confined near the clamps. The decay length scale is given by $l\sim \alpha^{-1}\sim T^{1/2}$ and is independent of the length of the film. The $l\sim T^{1/2}$ scaling of the penetration depth with tension indicates that as tension is increased keeping everything else constant, edge wrinkles become longer--as observed qualitatively in figure \ref{fig:fig_2} (f). Quantitative characterization of the decay length scale was carried out from FEA simulations on imperforate auxetic films. The results for the penetration length as a function of the applied tension are plotted on a log-log scale in figure \ref{fig:fig_3} (d) for five different values of $\nu < 0$. A family of lines with a slope 1/2 in the background provides the expected trend for the data, which is due to $T^{1/2}$ scaling. The $y$-intercept would depend on the film Poisson's ratio, indeed as observed here. Data measured from FEA simulations (figure \ref{fig:fig_3} (d)) show excellent alignment with the predicted trend. Since the actual stress field near the edge is complex and 2-D, not all wrinkles are exactly the same length--data for the longest wrinkle from simulations were used. 

\section{Conclusions}
In conclusion, laboratory experiments as well as computer simulations demonstrate that thin films can be stabilized within the bulk of a stretched film by the use of auxetic microstructuring in the bulk where wrinkles within non-auxetic films would otherwise appear. However, localized wrinkles appear at the clamps, as opposed to non-auxetic films in which clamped edges are locally taut. A simple theory is presented here, which is able to quantitatively characterize the anomalous edge wrinkles. Resistance to wrinkling, in conjunction with previously known synclastic curvature of auxetic film upon bending, make them a promising candidate for future applications such as artificial skin, cardiac patches and biomedical stretchable sensors. The phenomenon reported here could have interesting implications at smaller length scales, such as potential elastic stabilization of 2D materials and auxetic molecular sheets. Potential applications of edge wrinkling could be developed for force metrology, e.g. on the lines of  \cite{stafford2004buckling,huang2007capillary}, by calibrating tension in terms of edge wrinkle length.

 \appendix

\section{Experimental tests}
Thin lattice membranes were fabricated by engraving two different micro-patterns on uniform solid films to obtain lattice membranes with two different microstructures – one with positive apparent Poisson's ratio and one with negative apparent Poisson's ratio. The perforated samples were prepared using high precision laser cutting. A 0.102-mm-thick acetate film, commonly available in stationary stores for plain paper copier use, was attached firmly to a clean glass sheet. It was laser cut using the LS6840 LaserScript engraving machine that operates on CO2 laser tube technology. The cutting parameters—i.e. speed and power of the laser beam were optimized using a systematic study to obtain neat edges in the micro-patterns. The samples were manufactured by moving the laser beam at a speed of 63 mm/s and a laser tube power of 15 W. For the corners, a lower power of 13 W was chosen in order to obtain a clean cut with no burn into the polymer. The films were then attached using cyanoacrylates adhesives--commonly known as “Super Glue”--to wooden ends specifically designed to fit into the grips of the tensile testing machine later used to characterize the samples. The ends were also laser cut from 3.5-mm-thick wooden sheets using the settings suggested in the User's Manual by the manufacture LaserScript at a cutting speed of 20 mm/s and 35 W power.  The qualitative behavior of the auxetic membranes was studied at first by uniaxially stretch the films using a horizontal hand operated Tensometer make W Monsanto. Force-displacement curves were obtained by using the Instron electrical mechanical test machine. The acetate sheets were stretched uniaxially by $\Delta$, respectively equal to 1.6 mm for the first auxetic sample, 1.2 mm for the second auxetic sample and 3.6 mm for the sample with positive Poisson's ratio in the Instron tensile machine. Since no prior information was available about the stiffness of the samples, the calibration of the upper force limits (respectively 20N for the first auxetic sample, 12N for the second auxetic sample and 120N for the sample with positive Poisson's ratio) was performed initially by hand on trial samples. The test was performed in displacement control mode of the machine and the stretch was applied at a speed of 0.01 mm/s. Images of the wrinkling patterns (6 per sample) were recorded at constant displacement intervals (equal to $\Delta/6$ mm) using NIKON D3200 SLR camera, fixed on a tripod, using a Tamron AF 18-200 mm F/3.5-6.3 XR lens. The flash was synchronized with the camera and it has been positioned perpendicularly to the sample in order to reflect on the acetate surface and highlight the wrinkle profile. 

\section{Numerical simulations}
The geometry for laser cutting was imported as a solid structure from the commercial solid modeling software Rhinoceros 5 (Robert McNeel \& Associates) which was used for the manufacture of the perforated films as well as finite element structural analysis using the commercial code Abaqus 6.13. The thin film was modeled as a 3D structure using the element type C3D8I (8-node brick element) within Abaqus. The geometric non-linearity capability of this element was enabled during the simulations. The number of degrees-of-freedom per node is 3 and the total number of elements over 35,000 were required by systematically refining the mesh till convergence was achieved. Acetate sheets have a Young's modulus of 1.6 GPa and Poisson's ratio of 0.38 (MatBase, 2016) which have been used for the computational simulations. The film material had positive Poisson's ratio; the architecture gives the film in Figure \ref{fig:fig_1} (a)-(d), \ref{fig:fig_2} (c), (d), (f) a negative apparent Poisson's ratio.

%% References


\bibliographystyle{model1-num-names}
\bibliography{sample.bib}






% \end{thebibliography}

\end{document}